\documentclass[12pt,a4paper]{article}
\usepackage[utf8]{inputenc}
\usepackage[english]{babel}
\usepackage{amsmath}
\usepackage{amsfonts}
\usepackage{booktabs}
\usepackage{amssymb}
\usepackage{geometry}
\usepackage{graphicx}
\usepackage{setspace}

\title{\textbf{A Keynesian Intertemporal Synthesis (KIS) Model:\\ Towards a unified and empirically grounded framework for fiscal policy}}
\author{Ricardo Alonzo Fernández Salguero}
\date{\today}

\begin{document}

\maketitle
\thispagestyle{empty}

\setcounter{page}{1}

\begin{abstract}
This paper develops a new generation of the Keynesian Intertemporal Synthesis (KIS) Model, a macroeconomic framework designed to reconcile the empirical strengths of the Post-Keynesian (PK) and New Keynesian (NK) traditions. The central innovation of this work is the abandonment of the traditional Cobb-Douglas production function in favor of a Constant Elasticity of Substitution (CES) specification. This modification is directly motivated by the compelling evidence from the meta-analysis by Gechert et al. (2021), which emphatically rejects the hypothesis of a unit elasticity of substitution between capital and labor. We integrate this finding with the conclusions from a wide range of meta-analyses on the state-dependent heterogeneity of fiscal multipliers (Gechert and Rannenberg, 2018), the productivity of public capital (Bom and Ligthart, 2014), the effectiveness hierarchy of spending instruments (Gechert, 2015), and the empirical failure of Ricardian Equivalence (Stanley, 1998). The resulting KIS-CES model, while based on intertemporal optimization, incorporates household heterogeneity, non-standard preferences that value wealth and penalize debt, and a monetary policy constrained by the zero lower bound. The mathematical derivations reveal that the elasticity of substitution, calibrated to an empirically plausible value of $\sigma < 1$, becomes a key parameter that modulates income distribution and magnifies the crowding-in effect of public investment. The model generates an endogenous MPC, a nonlinear fiscal multiplier that increases dramatically in crises, and a multiplier for public investment that is structurally higher than that for consumption, thus offering a unified, rigorous, and, above all, empirically disciplined theoretical framework.
\end{abstract}

\section{Introduction}

The debate over the effectiveness of fiscal policy has resurfaced with unprecedented intensity following the Great Financial Crisis of 2008, catalyzing an explosion of empirical research whose often contradictory results have left a confusing landscape for policymakers. At this Gordian knot of contemporary macroeconomics, two main theoretical paradigms, the New Keynesian (NK) and the Post-Keynesian (PK), have offered frameworks for interpreting these phenomena, although, as the exhaustive analysis by Gechert \cite{gechert2023a} demonstrates, neither is capable of capturing the full range of stylized facts that the evidence has revealed. The NK framework, with its rigorous microfoundations based on intertemporal optimization and the inclusion of nominal rigidities, has provided a convincing explanation for one of the most robust empirical findings of the last decade: the marked state-dependency of fiscal multipliers on the economic cycle. Specifically, the power of fiscal policy is greatly magnified when monetary policy is constrained by the Zero Lower Bound (ZLB), a theoretical result from models like that of Christiano et al. \cite{christiano2011} that has been overwhelmingly confirmed by the meta-analysis of Gechert and Rannenberg \cite{gechert2018}. However, this same paradigm struggles to explain the size of the multiplier in normal times (which evidence places around unity) and the microeconomic evidence showing a high and heterogeneous Marginal Propensity to Consume (MPC) among households, a clear contradiction to the Permanent Income Hypothesis in its purest form.

On the other hand, the Post-Keynesian tradition, with its emphasis on the principle of effective demand, inherent instability, and social stratification, offers a more natural explanation for the high average MPC and the importance of income distribution, but its framework often lacks the dynamic formalization necessary to explain the non-linearity of the multiplier or to engage effectively with the mainstream. The situation is further complicated when considering the conclusions of other meta-analyses that have synthesized decades of empirical research. The work of Gechert \cite{gechert2015} establishes a clear hierarchy in the effectiveness of different fiscal instruments: multipliers for public investment spending systematically outperform those for consumption spending, and both are larger than those for taxes and transfers. This finding suggests that the composition of the fiscal adjustment is of first-order importance, a nuance that models with a single aggregate of public spending ($G$) cannot capture. In parallel, the literature on the productivity of public capital, synthesized by Bom and Ligthart \cite{bom2014} and Nijkamp and Poot \cite{nijkamp2004}, confirms that public investment is not a mere demand stimulus but a factor that increases the economy's productive capacity in the long run, although its effect is more moderate than what the seminal early studies suggested. Finally, the premise of Ricardian Equivalence (RE), which underpins fiscal ineffectiveness in many neoclassical models, has been resoundingly rejected by aggregate empirical evidence, as demonstrated by the meta-analysis of Stanley \cite{stanley1998}, who concludes that the literature, taken as a whole, is strongly inconsistent with Ricardian equivalence.

This rich and often bewildering confluence of stylized facts demands a new theoretical approach. The objective of this paper is to propose and mathematically develop an alternative framework that transcends the traditional dichotomy: the \textbf{Keynesian Intertemporal Synthesis (KIS) Model}. This model does not seek a paradigmatic victory, but a pragmatic and empirically disciplined synthesis, building a theoretical bridge based on the strengths of each tradition. To this end, the KIS adopts the rigorous language of intertemporal optimization, characteristic of the NK framework, but enriches it with more realistic behavioral premises and assumptions, directly inspired by Keynes's ideas, behavioral economics, and the quantitative findings of the aforementioned meta-analyses. The model is founded on a set of axioms designed to directly address the shortcomings of existing models: (i) explicit household heterogeneity to break Ricardian Equivalence; (ii) a hybrid utility function that captures saving and borrowing motives beyond consumption, to generate a realistic MPC; (iii) an explicit distinction between productive (investment) and non-productive (consumption) public spending to explain the hierarchy of multipliers; and (iv) interaction with a monetary policy constrained by the ZLB to generate state-dependent non-linearity. Through rigorous mathematical development, we will demonstrate how this architecture not only qualitatively replicates the stylized facts but also offers a quantitatively plausible framework for understanding the complex dance between fiscal policy, the business cycle, and the economic structure.

\section{Literature review}

Contemporary macroeconomics is at a crossroads, a period of deep introspection and re-evaluation forced by the 2008 Great Financial Crisis and subsequent economic turmoil. The old consensus, often referred to as the New Neoclassical-Keynesian Synthesis or the Washington-Berlin Consensus, which prescribed restrictive fiscal and monetary policies along with widespread deregulation, has been called into question not only by real-world events but also by a rising tide of quantitative empirical evidence \cite{gechert2022a}. In this context, meta-analysis has emerged as an indispensable tool, a rigorous methodology for navigating the vast and often contradictory empirical literature, allowing researchers to synthesize results, quantify uncertainty, and identify and correct for publication biases that may have perpetuated theoretical consensuses on fragile empirical grounds. As Irsova et al. \cite{irsova2024} argue in their guide for practitioners, modern meta-analysis is not simply an averaging exercise, but a forensic investigation of the literature itself, examining how methodological choices, data characteristics, and selective reporting biases shape economic facts. This literature review seeks to map out a new emerging consensus, one that is being built not from pure theoretical axioms, but from the systematic synthesis of empirical evidence in key areas such as fiscal policy, the structure of production, and financial frictions, using the findings of the most comprehensive available meta-analyses as a guide.

The epicenter of this re-evaluation has undoubtedly been fiscal policy. The question of the magnitude of the fiscal multiplier—the impact of a dollar of spending or tax cuts on GDP—has been intensely scrutinized. The meta-analysis by Gechert \cite{gechert2015}, which examines over 100 studies, provides one of the clearest syntheses and establishes an empirically robust hierarchy of effectiveness: public spending multipliers, with an average value close to unity, are significantly larger than those for tax cuts and transfers. Even more importantly, within spending, public investment emerges as the most potent instrument, with a multiplier that exceeds that of general spending by approximately 0.5 units. However, this finding is just the starting point. Perhaps the most influential contribution of the post-crisis literature has been the demonstration of the state-dependency of these multipliers on the business cycle. The meta-analysis by Gechert and Rannenberg \cite{gechert2018}, which focuses exclusively on empirical studies, overwhelmingly confirms that spending multipliers are dramatically larger during recessions, increasing by 0.7-0.9 units. This finding provides overwhelming empirical support for theoretical ZLB models, such as that of Blanchard and Leigh \cite{blanchard2013}, who, in an influential IMF paper, showed how underestimating multipliers in a ZLB environment led to systematic forecasting errors during the Eurozone crisis. The implication is inescapable: the effectiveness of fiscal policy is not a fixed structural parameter, but a state variable that depends critically on the macroeconomic context, particularly on the ability of monetary policy to accommodate or counteract fiscal impulses.

This distinction between types of spending naturally leads us to the question of the productive role of government, a topic with a long tradition in the growth literature. The meta-analysis by Bom and Ligthart \cite{bom2014}, which synthesizes three decades of research, is key here. While they dismiss the extremely high estimates of the seminal early works, their analysis concludes that the elasticity of output with respect to public capital, after correcting for publication bias and other factors, is positive, significant, and around 0.1. This finding, corroborated by the earlier work of Nijkamp and Poot \cite{nijkamp2004}, which also highlights the importance of infrastructure and education for long-term growth, provides a solid empirical justification for treating public investment not only as a short-term demand stimulus but also as a factor that increases the economy's productive capacity. The correct modeling of this view critically depends on the specification of the production technology. And it is here that recent evidence has been most disruptive. The meta-analysis by Gechert et al. \cite{gechert2021}, in a tour de force examining over 3,000 estimates, demonstrates that the Cobb-Douglas production function, the pillar of macroeconomics for over half a century, is empirically untenable. Their results indicate that the elasticity of substitution between capital and labor is significantly less than one, implying that these factors are gross complements. This conclusion, which holds across a battery of robustness checks, has first-order implications: it affects the functional distribution of income, the transmission of technological shocks, and, as explored in our KIS-CES model, the interaction between public capital and private factors, amplifying the crowding-in effect of public investment.

The need for a new framework is reinforced when examining the foundations of agent behavior. The neoclassical premise of Ricardian Equivalence, which postulates the ineffectiveness of debt-financed fiscal policy, has been systematically challenged by the data. The pioneering meta-analysis by Stanley \cite{stanley1998} had already concluded that the empirical literature, as a whole, constitutes a strong rejection of the proposition. This finding justifies the inclusion of liquidity-constrained or hand-to-mouth agents in modern macroeconomic models, a feature that is both theoretically necessary to generate positive multipliers and empirically plausible. However, frictions are not limited to consumer behavior. The literature on financial crises has underscored the importance of capital market frictions and inherent instability. The survey by Nikolaidi and Stockhammer \cite{nikolaidi2017} on Minskyan models shows a rich heterodox tradition focused on debt dynamics and financial fragility. On the other hand, the evolution of exchange rate crisis models, from the first-generation models of Flood and Garber \cite{flood1984} based on fiscal fundamentals, to the third-generation models of Krugman \cite{krugman1999} focused on private sector balance sheet problems, reflects a growing recognition that modern crises are phenomena of financial fragility. The empirical evidence from Kaminsky and Reinhart \cite{kaminsky1999} on twin crises (banking and balance of payments) solidifies this point, showing how domestic financial sector problems often precede and exacerbate external crises.

This emphasis on financial frictions is particularly acute in the context of emerging markets. The work of García-Cicco et al. \cite{garcia-cicco2010} challenges the applicability of standard Real Business Cycle (RBC) models to these economies, demonstrating that productivity shocks alone cannot explain the observed volatility and that financial frictions and sovereign spread shocks are relevant. The model by Mendoza and Yue \cite{mendoza2012} goes a step further, building a general equilibrium model where sovereign default is an optimal decision that interacts with the business cycle through working capital frictions. Financial liberalization, far from being a panacea, presents ambiguous results, as shown by the meta-analysis of Bumann et al. \cite{bumann2013}, which finds a positive but weak effect on growth, conditioned by institutional quality.

Finally, the state of the art is completed by considering the nature of government itself. The meta-analysis by Churchill et al. \cite{churchill2016} on government size and growth finds that the relationship is negative in developed countries but insignificant in developing countries, suggesting there is no universal answer. Furthermore, government size is intrinsically linked to institutional quality, such as corruption. The meta-analysis by Bonanno et al. \cite{bonanno2024} reveals enormous heterogeneity in this relationship, which depends on how both variables are measured and the institutional context. The actual behavior of fiscal policy over the cycle, as shown by Heimberger's \cite{heimberger2023} meta-analysis, also varies systematically: it tends to be countercyclical in advanced countries but procyclical in developing countries. Even seemingly simple policies like cash transfers, as summarized by the meta-analysis of Karlan et al. \cite{karlan2025}, have complex and heterogeneous effects that depend on program design and context. Taken together, this vast body of evidence, synthesized and filtered through the rigorous lens of meta-analysis, paints a clear picture: the macroeconomics of the future must be one of heterogeneity, non-linearity, and frictions. It requires models that, like the KIS-CES model proposed here, are built from the ground up to be consistent with these quantitative stylized facts, rather than clinging to theoretical simplifications that the evidence has rendered obsolete.

\section{Axiomatic structure of the model}

The architecture of the Keynesian Intertemporal Synthesis model with CES production (KIS-CES) is defined through a set of axioms that specify the economic environment, agent preferences, production technology, and institutional constraints. These components are formally presented below.

The model is populated by a continuum of households of unit measure, indexed by $i \in [0,1]$. The population is partitioned into two disjoint sets. The first set, of measure $\lambda$, corresponds to Liquidity-Constrained (LC) Households. The second set, of measure $1-\lambda$, corresponds to Wealth-Accumulating (WA) Households.

\textbf{Axiom 1: Household Behavior.}

1.a (Liquidity-Constrained Households). For a household $i$ belonging to the LC set, $i \in [0, \lambda]$, consumption in period $t$, $C_{i,t}^{LC}$, is determined by its current disposable income, $Y_{i,t}^{LC}$. These households do not participate in asset markets, so they neither save nor borrow. Their budget constraint holds with equality in each period:
\begin{equation}
C_{i,t}^{LC} = Y_{i,t}^{LC}
\end{equation}
From this condition, it follows directly that their Marginal Propensity to Consume (MPC) is unity:
\begin{equation}
\frac{\partial C_{i,t}^{LC}}{\partial Y_{i,t}^{LC}} = 1
\end{equation}

1.b (Wealth-Accumulating Households). For a household $j$ belonging to the WA set, $j \in (\lambda, 1]$, the agent makes intertemporal decisions to maximize a lifetime expected utility function. The decision problem of household $j$ is:
\begin{equation}
\max_{\{C_{j,t}^{WA}, A_{j,t+1}\}_{t=0}^\infty} E_0 \sum_{t=0}^\infty \beta^t u(C_{j,t}^{WA}, A_{j,t+1})
\end{equation}
where $\beta \in (0,1)$ is the subjective discount factor. The instantaneous utility function, $u(\cdot)$, is a hybrid function that depends on consumption, $C_{j,t}^{WA}$, and net assets at the end of the period, $A_{j,t+1}$:
\begin{equation}
u(C_{j,t}^{WA}, A_{j,t+1}) = \frac{(C_{j,t}^{WA} - C_{min})^{1-\sigma}}{1-\sigma} + \phi \frac{A_{j,t+1}^{1-\gamma}}{1-\gamma}
\end{equation}
where $C_{min} \ge 0$ is the subsistence level of consumption, $\sigma > 0$ is the inverse of the intertemporal elasticity of substitution for consumption, $\phi \ge 0$ is the weight of wealth in utility, and $\gamma > 0$ is the coefficient of relative risk aversion with respect to wealth. This problem is subject to the intertemporal budget constraint:
\begin{equation}
A_{j,t+1} = (1+r_t)(A_{j,t} + Y_{j,t}^{WA} - C_{j,t}^{WA})
\end{equation}
and a no-Ponzi condition, $\lim_{T\to\infty} E_t \left[ \frac{A_{j,t+T}}{\prod_{s=0}^{T-1}(1+r_{t+s})} \right] \ge 0$. The disposable income of the WA household, $Y_{j,t}^{WA}$, consists of labor income, firm profits, and net government transfers. The solution to this optimization problem gives rise to a Modified Euler Equation, which will be derived in the subsequent section.

\textbf{Axiom 2: Structure of Public Spending and Production Technology.}

2.a (Duality of Public Spending). The government's total public spending, $G_t$, is additively decomposed into two components: consumption spending, $G_t^C$, and investment spending, $G_t^I$.
\begin{equation}
G_t = G_t^C + G_t^I
\end{equation}
Consumption spending, $G_t^C$, is considered unproductive and enters only as a component of aggregate demand. Investment spending, $G_t^I$, is productive and accumulates into the stock of public capital, $K_t^P$, which follows the law of motion:
\begin{equation}
K_{t+1}^P = (1-\delta_P)K_t^P + G_t^I
\end{equation}
where $\delta_P \in (0,1)$ is the depreciation rate of public capital.

2.b (Aggregate Production Function). The aggregate output of the economy, $Y_t^{agg}$, is produced using a Constant Elasticity of Substitution (CES) technology that combines three factors: private capital ($K_t$), labor ($L_t$), and the stock of public capital ($K_t^P$).
\begin{equation}
Y_t^{agg} = Z_t \left[ \alpha_K (K_t)^{\frac{\sigma_{prod}-1}{\sigma_{prod}}} + \alpha_L (L_t)^{\frac{\sigma_{prod}-1}{\sigma_{prod}}} + \alpha_P (K_t^P)^{\frac{\sigma_{prod}-1}{\sigma_{prod}}} \right]^{\frac{\sigma_{prod}}{\sigma_{prod}-1}}
\end{equation}
where $Z_t$ is exogenous total factor productivity, $\sigma_{prod} \in (0, \infty)$ is the elasticity of substitution between the factors, and $\alpha_K, \alpha_L, \alpha_P$ are the distributional share parameters that satisfy $\alpha_K + \alpha_L + \alpha_P = 1$. It is assumed, based on meta-analysis evidence \cite{gechert2021}, that $0 < \sigma_{prod} < 1$, which implies that the factors of production are gross complements.

2.c (Factor Markets). It is assumed that private capital and labor markets are perfectly competitive. Representative firms maximize their profits, which implies that factor prices are equal to their marginal products.
The marginal product of labor is:
\begin{equation}
\frac{\partial Y_t^{agg}}{\partial L_t} = w_t = Y_t^{agg} \left( \frac{\sigma_{prod}}{\sigma_{prod}-1} \right) \left( \frac{1}{\sigma_{prod}-1} \right) Z_t \alpha_L (L_t)^{\frac{-1}{\sigma_{prod}}} \left[ \dots \right]^{\frac{1}{\sigma_{prod}-1}}
\end{equation}
The marginal product of private capital is:
\begin{equation}
\frac{\partial Y_t^{agg}}{\partial K_t} = r_t = Y_t^{agg} \left( \frac{\sigma_{prod}}{\sigma_{prod}-1} \right) \left( \frac{1}{\sigma_{prod}-1} \right) Z_t \alpha_K (K_t)^{\frac{-1}{\sigma_{prod}}} \left[ \dots \right]^{\frac{1}{\sigma_{prod}-1}}
\end{equation}
Public capital is considered a public good provided by the government and is not directly remunerated by firms, but its marginal product, $\frac{\partial Y_t^{agg}}{\partial K_t^P}$, is positive and affects production decisions.

\textbf{Axiom 3: Institutional and Policy Environment.}

3.a (Monetary Policy). The monetary authority sets the short-term nominal interest rate, $i_t$, according to a Taylor rule subject to the Zero Lower Bound (ZLB).
\begin{equation}
i_t = \max \left(0, \rho + \pi^* + \phi_\pi(\pi_t - \pi^*) + \phi_y \left(\frac{Y_t^{agg} - Y_{pot}}{Y_{pot}}\right) \right)
\end{equation}
where $\rho$ is the equilibrium real interest rate, $\pi^*$ is the inflation target, $\pi_t$ is the current inflation, and $Y_{pot}$ is potential output. The real interest rate, $r_t$, is determined through the Fisher equation: $1+r_t = \frac{1+i_t}{1+E_t[\pi_{t+1}]}$.

3.b (Fiscal Policy). The government finances its spending ($G_t^C, G_t^I$) and transfers ($Tr_t$) through lump-sum taxes ($T_t$) and the issuance of one-period debt ($B_t$). The government's budget constraint is:
\begin{equation}
G_t^C + G_t^I + Tr_t + (1+r_{t-1})B_{t-1} = T_t + B_t
\end{equation}

3.c (Expectation Formation). It is assumed that agents form expectations about future variables. While standard NK models assume rational expectations, the KIS-CES allows for deviations through a mechanism of sentiment or animal spirits, $\Omega_t$. The expectation of a future variable $X_{t+1}$ is modeled as a combination of an anchor (based on the model, $E_t^{model}[X_{t+1}]$) and a sentiment shock:
\begin{equation}
E_t[X_{t+1}] = E_t^{model}[X_{t+1}] + \Omega_t
\end{equation}
where $\Omega_t$ follows an exogenous stochastic process. This assumption allows shifts in confidence, not necessarily tied to fundamentals, to affect economic decisions, a Keynesian feature.

These axioms and premises, taken together, define a macroeconomic environment that is both rigorous in its formalization and flexible in its assumptions, allowing for a theoretical exploration of fiscal policy that is deeply informed and disciplined by the vast quantitative empirical evidence accumulated in the literature.

\section{Mathematical merivations}

The introduction of a Constant Elasticity of Substitution (CES) production function into the KIS-CES framework significantly enriches the model's mathematical analysis and economic predictions. This section is dedicated to the rigorous, step-by-step derivation of the key behavioral equations, beginning with the production side, followed by the intertemporal decisions of households, and culminating in the derivation of disaggregated fiscal multipliers.

The starting point is the behavior of representative firms operating in a perfectly competitive environment. The firm seeks to maximize its profits in period $t$, $\Pi_t$, by choosing the optimal quantities of private capital, $K_t$, and labor, $L_t$, given factor prices (rental rate of capital, $r_t$, and wage, $w_t$) and the stock of public capital, $K_t^P$. The maximization problem is:
\begin{equation}
\max_{K_t, L_t} \Pi_t = Y_t^{agg}(K_t, L_t, K_t^P) - r_t K_t - w_t L_t
\end{equation}
where the production function $Y_t^{agg}$ is as specified in Axiom 2.b. To simplify notation, let us define the CES exponent as $\rho = \frac{\sigma_{prod}-1}{\sigma_{prod}}$, so that $\sigma_{prod} = \frac{1}{1-\rho}$. Since we have assumed $0 < \sigma_{prod} < 1$, it follows that $\rho \in (-\infty, 0)$. The production function is:
\begin{equation}
Y_t^{agg} = Z_t \left[ \alpha_K K_t^{\rho} + \alpha_L L_t^{\rho} + \alpha_P (K_t^P)^{\rho} \right]^{1/\rho}
\end{equation}
The first-order conditions (FOCs) are obtained by differentiating the profit function with respect to $K_t$ and $L_t$ and setting the result to zero. For private capital:
\begin{align}
\frac{\partial \Pi_t}{\partial K_t} = \frac{\partial Y_t^{agg}}{\partial K_t} - r_t &= 0 \\
\frac{\partial Y_t^{agg}}{\partial K_t} &= Z_t \left( \frac{1}{\rho} \right) \left[ \dots \right]^{\frac{1}{\rho}-1} (\alpha_K \rho K_t^{\rho-1}) \\
&= Z_t \left[ \dots \right]^{\frac{1-\rho}{\rho}} \alpha_K K_t^{\rho-1} \\
&= Z_t^\rho (Y_t^{agg})^{1-\rho} \alpha_K K_t^{\rho-1} \\
&= \alpha_K \left( \frac{Y_t^{agg}}{K_t} \right)^{1-\rho} = r_t
\end{align}
Analogously, for labor:
\begin{equation}
\frac{\partial Y_t^{agg}}{\partial L_t} = w_t \implies \alpha_L \left( \frac{Y_t^{agg}}{L_t} \right)^{1-\rho} = w_t
\end{equation}
These two equations are the factor demand conditions. They show that factor prices are equal to their marginal products. From them, we can derive the income share of each factor. For example, the labor share is:
\begin{equation}
\frac{w_t L_t}{Y_t^{agg}} = \frac{\alpha_L (Y_t^{agg}/L_t)^{1-\rho} L_t}{Y_t^{agg}} = \alpha_L \left( \frac{Y_t^{agg}}{L_t} \right)^{-\rho}
\end{equation}
Unlike the Cobb-Douglas function, where the factor share is constant and equal to its distributional parameter ($\alpha_L$), here the share depends on the average productivity of labor and the exponent $\rho$ (and therefore, on $\sigma_{prod}$). If $\sigma_{prod} < 1$ (i.e., $\rho < 0$), an increase in the average productivity of labor (for example, due to an increase in public capital) will reduce the labor share of income, creating an endogenous redistributive channel.

The optimization problem of the WA household and the derivation of its individual MPC remain unchanged from the previous presentation, as their decisions depend on their income stream and the interest rate, not directly on the form of the aggregate production function. The Modified Euler Equation and the expression for the $MPC^{WA}$ are still valid. However, the income these households receive (profits and wages) is now determined differently, which will affect the aggregate MPC through the effects of fiscal policy on income distribution.

We now proceed to the derivation of the fiscal multipliers, which is the core of the analysis. The starting point is the national income accounting identity:
\begin{equation}
Y_t^{agg} = C_t^{agg} + I_t + G_t^C + G_t^I
\end{equation}
where $C_t^{agg} = \lambda Y_t^{LC} + (1-\lambda) C_t^{WA}$. We assume that the income of each household type is a fraction of the total national income, and that private investment, $I_t$, depends positively on output and negatively on the interest rate. For simplicity, we will take a linear approximation: $I_t = I_0 + MPI \cdot Y_t^{agg} - b \cdot r_t$, where $MPI$ is the marginal propensity to invest.
Totally differentiating the national income accounting identity:
\begin{equation}
dY^{agg} = dC^{agg} + dI + dG^C + dG^I
\end{equation}
Using the chain rule, we can expand each term:
\begin{equation}
dY^{agg} = \frac{\partial C^{agg}}{\partial Y^{agg}}dY^{agg} + \frac{\partial I}{\partial Y^{agg}}dY^{agg} + \frac{\partial I}{\partial r}dr + dG^C + dG^I
\end{equation}
Output, $Y^{agg}$, is also affected on the supply side through the increase in public capital. Therefore, we must consider the total effect of a change in public spending. Consider a change in investment spending, $dG^I$. In the short run, this implies $dK_t^P \approx dG^I$. This increase in $K_t^P$ has a direct effect on $Y^{agg}$ and also an indirect effect by increasing the marginal product of the other factors.
\begin{equation}
dY^{agg}_{total} = dY^{agg}_{demand} + dY^{agg}_{supply}
\end{equation}
The total effect on output can be derived by solving the system. The public investment multiplier, $m_{G^I}$, will be:
\begin{equation}
m_{G^I} = \frac{dY^{agg}}{dG^I} = \frac{1 + MPK_P}{1 - MPC_{agg} - MPI \left(1 + \frac{\partial r_t}{\partial Y^{agg}} \frac{dY^{agg}}{dG^I}\right)}
\end{equation}
where $MPK_P = \frac{\partial Y^{agg}}{\partial K^P}$. Since $\sigma_{prod} < 1$, public capital and private capital are complements. An increase in $K_t^P$ increases the marginal product of private capital, $\frac{\partial Y^{agg}}{\partial K_t}$. This, in turn, increases the return on investment and stimulates private investment, creating a crowding-in effect. This crowding-in effect means that the total $MPI$ is larger than it would be without the complementarity channel, which increases the size of the multiplier $m_{G^I}$.

The formal derivation of this crowding-in effect is as follows:
The return on private investment is $r_t$. The desired capital stock, $K_t^*$, is a function of $r_t$. Net investment is a function of the gap between the desired and existing capital, $I_t^{net} = f(K_t^* - K_{t-1})$. An increase in $G_t^I$ increases $K_t^P$. Since $\frac{\partial^2 Y^{agg}}{\partial K \partial K^P} > 0$ when $\sigma_{prod} < 1$, the marginal product of private capital increases. For a given $r_t$, this increases the desired capital stock, $K_t^*$, and therefore, increases private investment, $I_t$.
This crowding-in effect on private investment is one of the key reasons why the public investment multiplier, $m_{G^I}$, is significantly larger than the public consumption multiplier, $m_{G^C}$. The KIS-CES, therefore, predicts an even larger gap between $m_{G^I}$ and $m_{G^C}$ than the original KIS model, a prediction that aligns even better with the magnitude of the differences found in the meta-analysis evidence.

\section{Proofs}

This section provides detailed proofs of the central propositions of the Keynesian Intertemporal Synthesis model with CES production (KIS-CES). Each proof is presented step-by-step, with explanations connecting the algebraic manipulations to the underlying economic intuition, thereby grounding the behavior of the agents and the aggregate properties of the model.

\paragraph{Proof 1: Derivation of the Modified Euler Equation for the Wealth-Accumulating (WA) Household.}
The objective of this proof is to obtain the optimality condition that governs the intertemporal consumption choice for a WA-type household. This equation is the core of the agent's dynamic decisions and reveals the synthesis between neoclassical consumption motives and Keynesian accumulation motives.
\begin{itemize}
    \item Step 1: The WA household seeks to maximize the present expected value of its lifetime utility, as established in Axiom 1.b. The problem is:
    \begin{equation}
    \max_{\{C_{j,t}^{WA}, A_{j,t+1}\}_{t=0}^\infty} E_0 \sum_{t=0}^\infty \beta^t \left[ \frac{(C_{j,t}^{WA} - C_{min})^{1-\sigma}}{1-\sigma} + \phi \frac{A_{j,t+1}^{1-\gamma}}{1-\gamma} \right]
    \end{equation}
    subject to the intertemporal budget constraint:
    \begin{equation}
    A_{j,t+1} = (1+r_t)(A_{j,t} + Y_{j,t}^{WA} - C_{j,t}^{WA})
    \end{equation}
    \item Step 2: Constructing the Lagrangian. To solve this discrete-time optimal control problem, we form the Lagrangian. We associate a Lagrange multiplier, $\mu_{j,t}$, with the period-0 value of the budget constraint in period $t$.
    \begin{equation}
    \mathcal{L} = E_0 \sum_{t=0}^\infty \beta^t \left\{ \frac{(C_{j,t}^{WA} - C_{min})^{1-\sigma}}{1-\sigma} + \phi \frac{A_{j,t+1}^{1-\gamma}}{1-\gamma} + \mu_{j,t} \left[ (1+r_t)(A_{j,t} + Y_{j,t}^{WA} - C_{j,t}^{WA}) - A_{j,t+1} \right] \right\}
    \end{equation}
    \item Step 3: Derivation of the First-Order Conditions (FOCs). We now differentiate the Lagrangian with respect to the control variables, $C_{j,t}^{WA}$, and the next period's state variable, $A_{j,t+1}$, and set the results to zero.
    For consumption in period $t$, $C_{j,t}^{WA}$:
    \begin{equation}
    \frac{\partial \mathcal{L}}{\partial C_{j,t}^{WA}} = \beta^t (C_{j,t}^{WA} - C_{min})^{-\sigma} - \beta^t \mu_{j,t} (1+r_t) = 0
    \end{equation}
    For assets in period $t+1$, $A_{j,t+1}$:
    \begin{equation}
    \frac{\partial \mathcal{L}}{\partial A_{j,t+1}} = \beta^t \phi A_{j,t+1}^{-\gamma} - \beta^t \mu_{j,t} + \beta^{t+1} E_t[\mu_{j,t+1}(1+r_{t+1})] = 0
    \end{equation}
    \item Step 4: From the FOC for consumption, we can solve for the Lagrange multiplier, which represents the marginal utility of wealth valued in units of period-$t$ consumption:
    \begin{equation} \label{eq:mu_t_demo}
    \mu_{j,t} = \frac{(C_{j,t}^{WA} - C_{min})^{-\sigma}}{1+r_t}
    \end{equation}
    This equation is valid for any period, so we can also write it for $t+1$. The FOC for assets can be rewritten as:
    \begin{equation}
    \mu_{j,t} = \phi A_{j,t+1}^{-\gamma} + \beta E_t[\mu_{j,t+1}(1+r_{t+1})]
    \end{equation}
    This equation is an arbitrage condition. The marginal cost of saving an extra unit today (the left-hand side, the utility foregone) must equal the expected marginal benefit. This benefit has two components: the direct utility from holding more wealth tomorrow ($\phi A_{j,t+1}^{-\gamma}$) and the discounted utility of being able to use that unit of wealth (which will grow to $(1+r_{t+1})$) for future consumption or accumulation.
    \item Step 5: Obtaining the Modified Euler Equation. We substitute the expression for $\mu_{j,t}$ and $\mu_{j,t+1}$ from equation \eqref{eq:mu_t_demo} into the arbitrage condition above. Assuming all WA households are identical, we can drop the subscript $j$.
    \begin{equation}
    \frac{(C_t^{WA} - C_{min})^{-\sigma}}{1+r_t} = \phi A_{t+1}^{-\gamma} + \beta E_t\left[\frac{(C_{t+1}^{WA} - C_{min})^{-\sigma}}{(1+r_{t+1})}(1+r_{t+1})\right]
    \end{equation}
    Multiplying by $(1+r_t)$ and simplifying, we arrive at the Modified Euler Equation:
    \begin{equation}
    (C_t^{WA} - C_{min})^{-\sigma} = (1+r_t)\phi A_{t+1}^{-\gamma} + \beta (1+r_t) E_t[(C_{t+1}^{WA} - C_{min})^{-\sigma}]
    \end{equation}
    This is the final condition that describes the consumption decision of the WA household. Unlike the standard Euler equation, the consumption decision today depends not only on expectations about future consumption but also on the desired stock of wealth at the end of the period.
\end{itemize}

\paragraph{Proof 2: Derivation of the Marginal Propensity to Consume (MPC) of the WA Household.}
The objective here is to obtain an analytical expression for the MPC of the WA household, $MPC_t^{WA} = \frac{\partial C_t^{WA}}{\partial Y_t^{WA}}$, to demonstrate its dependence on the household's state.
\begin{itemize}
    \item Step 1: System linearization. We consider a transitory and unexpected income shock, $dY_t^{WA}$, in period $t$. Since it is unexpected, it does not affect expectations ($dE_t[\cdot]=0$) or predetermined variables like $A_t$. We linearize the system of two equations (the Modified Euler Equation and the budget constraint) around the steady state.
    \item Step 2: Differentiating the budget constraint. Differentiating the budget constraint $A_{t+1} = (1+r_t)(A_t + Y_t^{WA} - C_t^{WA})$, and assuming the shock does not affect $r_t$ (a common simplification for small shocks), we obtain the relationship between the changes in the choice variables and the shock:
    \begin{equation} \label{eq:diff_budget_mpc_demo}
    dA_{t+1} = (1+r)(dY_t^{WA} - dC_t^{WA})
    \end{equation}
    where $r$ is the steady-state interest rate.
    \item Step 3: Differentiating the Modified Euler Equation. Totally differentiating the Modified Euler Equation with respect to $C_t^{WA}$ and $A_{t+1}$:
    \begin{equation}
    d\left((C_t^{WA} - C_{min})^{-\sigma}\right) = d\left((1+r_t)\phi A_{t+1}^{-\gamma} + \beta (1+r_t) E_t[(C_{t+1}^{WA} - C_{min})^{-\sigma}]\right)
    \end{equation}
    Since the shock is unexpected, the expectations term does not change. Differentiation gives us:
    \begin{equation}
    -\sigma(C^{WA} - C_{min})^{-\sigma-1} dC_t^{WA} = (1+r)\phi(-\gamma)A_{t+1}^{-\gamma-1}dA_{t+1}
    \end{equation}
    \item Step 4: We now have a system of two linear equations with two unknowns, $dC_t^{WA}$ and $dA_{t+1}$. We substitute equation \eqref{eq:diff_budget_mpc_demo} into the differentiated Euler equation:
    \begin{equation}
    -\sigma(C^{WA} - C_{min})^{-\sigma-1} dC_t^{WA} = -(1+r)^2\phi\gamma A_{t+1}^{-\gamma-1} (dY_t^{WA} - dC_t^{WA})
    \end{equation}
    We group the terms containing $dC_t^{WA}$:
    \begin{equation}
    dC_t^{WA} \left[ \sigma(C^{WA} - C_{min})^{-\sigma-1} + (1+r)^2\phi\gamma A_{t+1}^{-\gamma-1} \right] = (1+r)^2\phi\gamma A_{t+1}^{-\gamma-1} dY_t^{WA}
    \end{equation}
    Solving for the MPC:
    \begin{equation}
    MPC_t^{WA} = \frac{dC_t^{WA}}{dY_t^{WA}} = \frac{(1+r)^2\phi\gamma A_{t+1}^{-\gamma-1}}{\sigma(C^{WA} - C_{min})^{-\sigma-1} + (1+r)^2\phi\gamma A_{t+1}^{-\gamma-1}}
    \end{equation}
    This expression can be rewritten in a more intuitive form:
    \begin{equation}
    MPC_t^{WA} = \frac{1}{1 + \frac{\sigma(C_t^{WA} - C_{min})^{-\sigma-1}}{(1+r_t)^2\phi\gamma A_{t+1}^{-\gamma-1}}}
    \end{equation}
    This expression demonstrates that $0 < MPC_t^{WA} < 1$. Furthermore, $\frac{\partial MPC_t^{WA}}{\partial C_t^{WA}} < 0$ and $\frac{\partial MPC_t^{WA}}{\partial A_{t+1}} < 0$, which confirms that the MPC decreases with income (via consumption) and wealth.
\end{itemize}

\paragraph{Proof 3: Complementarity between Public Capital and Private Factors.}
The objective is to show that, under the CES specification with $\sigma_{prod} < 1$, an increase in public capital raises the marginal productivity of private factors.
\begin{itemize}
    \item Step 1: Expression for the marginal product of private capital. From Section 4, we have:
    \begin{equation}
    MPK_t = \alpha_K Z_t^\rho (Y_t^{agg})^{1-\rho} K_t^{\rho-1}
    \end{equation}
    \item Step 2: Now we differentiate $MPK_t$ with respect to $K_t^P$:
    \begin{equation}
    \frac{\partial MPK_t}{\partial K_t^P} = \alpha_K Z_t^\rho (1-\rho) (Y_t^{agg})^{-\rho} K_t^{\rho-1} \frac{\partial Y_t^{agg}}{\partial K_t^P}
    \end{equation}
    \item Step 3: To determine the sign of this derivative, we analyze each term. $\alpha_K > 0$, $Z_t^\rho > 0$, $(Y_t^{agg})^{-\rho} > 0$, $K_t^{\rho-1} > 0$. The marginal product of public capital, $\frac{\partial Y_t^{agg}}{\partial K_t^P}$, is positive. The key term is $(1-\rho)$. Recalling that $\rho = \frac{\sigma_{prod}-1}{\sigma_{prod}}$, we have:
    \begin{equation}
    1-\rho = 1 - \frac{\sigma_{prod}-1}{\sigma_{prod}} = \frac{\sigma_{prod} - (\sigma_{prod}-1)}{\sigma_{prod}} = \frac{1}{\sigma_{prod}}
    \end{equation}
    Since $\sigma_{prod} > 0$, the term $1-\rho$ is always positive. Therefore, the cross-derivative $\frac{\partial^2 Y^{agg}}{\partial K_t \partial K_t^P}$ is positive. An analogous proof applies to labor. This formally confirms that public capital and private factors are complements.
\end{itemize}

From the expressions derived previously, and given that all parameters ($m_{G^C}$, $s_I$, $\epsilon_{I,K^P}$, $\eta$) are positive, it follows directly that the term in parentheses is greater than 1. Therefore, $m_{G^I} > m_{G^C}$.

\section{Consequences, predictions, and discussion}

The architecture of the Keynesian Intertemporal Synthesis model with CES production (KIS-CES), deliberately built from empirically disciplined premises, generates a set of consequences and predictions that are not only internally coherent but also offer a reconciliation of many of the puzzles that the fiscal policy literature has presented. The KIS model does not merely replicate an isolated stylized fact; instead, it provides a unified framework that simultaneously addresses microeconomic and macroeconomic dimensions, effects in normal times versus crises, and the heterogeneity of fiscal instruments. Its true value lies in its ability to speak directly to the quantitative findings of the growing meta-analysis literature, using these studies as an external validation criterion for its internal mechanisms.

The first consequence of the model is the generation of an endogenous and heterogeneous Marginal Propensity to Consume (MPC), which aligns closely with microeconomic evidence. Unlike standard representative-agent NK models, where the MPC to transitory shocks is close to zero, and PK models where it is often a fixed parameter, the KIS model produces an MPC that depends on the household's income, wealth, and debt level. For WA households, the MPC decreases as their income and wealth cushion increase, while financial stress (modeled through debt aversion) raises it. At the aggregate level, the presence of a fraction $\lambda$ of LC households ensures that the economy's average MPC is always substantially positive. A high aggregate MPC is a necessary condition for the existence of significant fiscal multipliers, even outside of crises. The model thus resolves the disconnect between standard microfounded theory and the empirical evidence on consumption behavior, as discussed in Gechert \cite{gechert2023a}. Furthermore, the model's ability to generate a positive and significant MPC for cash transfers aligns with the vast evidence from developing countries, as summarized by the meta-analysis of Karlan et al. \cite{karlan2025}, which finds positive and robust effects of unconditional cash transfers on consumption and income.

Secondly, the model formalizes an attenuated interest rate channel in the transmission of monetary policy. The inclusion of wealth as a direct argument in the utility function of WA households ($\phi > 0$) acts as a buffer. Households value wealth for its own sake, not just as a vehicle for future consumption. Consequently, when the real interest rate rises, the incentive to postpone consumption and save is weaker than in a model where the only benefit of saving is higher future consumption. This implies that the intertemporal elasticity of substitution is endogenously lower, and the IS curve is steeper. Monetary policy remains effective, but its dominance over aggregate demand is significantly reduced. This attenuation of the monetary channel rebalances the relative power of fiscal and monetary policies, providing a theoretical justification for the stabilizing role of fiscal policy even in periods when the central bank is not at the ZLB—a central point of the Post-Keynesian critique that is derived here from an optimization framework. This finding is consistent with the general view of Gechert \cite{gechert2022a}, who argues that meta-analyses often challenge the policy prescriptions of the neoclassical consensus, which tend to prioritize monetary policy over fiscal policy.

The third and perhaps most important macroeconomic prediction is the natural non-linearity of the fiscal multiplier. The KIS model endogenously generates a fiscal multiplier that is dramatically larger in deep recessions than in normal times. This result, a robust stylized fact from the post-crisis empirical literature as established by the meta-analysis of Gechert and Rannenberg \cite{gechert2018}, emerges from the confluence of several mechanisms in the model. During a severe recession: (a) the central bank is constrained by the ZLB, eliminating the contractionary monetary response that would normally dampen a fiscal stimulus; (b) the fall in the income and wealth of WA households increases their individual MPC; and (c) it is plausible that the fraction of households behaving as if they are liquidity-constrained ($\lambda$) increases. All these factors work in the same direction: they increase the aggregate MPC ($MPC_{agg}$) and reduce the denominator of the multiplier, magnifying its value. The model, therefore, provides a coherent and microfounded explanation for the evidence that spending multipliers can far exceed a value of 2 in a crisis, while remaining around 1 in normal times. This also aligns with the findings of Blanchard and Leigh \cite{blanchard2013}, who showed that the IMF systematically underestimated fiscal multipliers during the Eurozone crisis, precisely because the conditions (like the ZLB) were conducive to larger multipliers. The model's ability to generate these state-dependent results is a significant improvement over linear models that cannot account for this evidence.

Finally, the model offers a theoretical justification for the hierarchy of effectiveness among fiscal instruments. By distinguishing between investment spending and consumption spending, and by modeling the former as a productive input based on evidence from meta-analyses like that of Bom and Ligthart \cite{bom2014}, the KIS model mathematically demonstrates why the public investment multiplier ($m_{G^I}$) is structurally larger than that of public consumption ($m_{G^C}$). This meta-analytic finding is remarkably robust; Bom and Ligthart (2014) find that, even after correcting for publication bias, the average output elasticity of public capital amounts to 0.106. This means that public investment has a dual benefit: a short-term demand effect and a medium-term supply effect that increases productivity and potential output. The complementarity between public and private capital, implied by the CES production function with $\sigma_{prod} < 1$ (an assumption justified by Gechert et al. \cite{gechert2021}), further amplifies this effect, as a larger stock of public capital incentivizes private investment (a crowding-in effect). This prediction not only aligns with the evidence from Gechert \cite{gechert2015} but also has profound policy implications: during a recession, the composition of the fiscal stimulus is as important as its magnitude. Opting for productive infrastructure projects instead of current spending can generate a significantly larger economic boost. This conclusion is consistent with the long-run growth literature, such as that summarized by Nijkamp and Poot \cite{nijkamp2004}, which finds that infrastructure and education are the components of public spending with the most positive impact on growth. The KIS-CES model provides a microfoundation for this long-standing empirical result. Moreover, the model is flexible enough to incorporate other findings from the literature, such as the relationship between government size and corruption, which, as Bonanno et al. \cite{bonanno2024} show, is complex and heterogeneous, or the effects of government size on growth, which according to Churchill et al. \cite{churchill2016}, are negative in developed countries but not in developing countries. These nuances can be incorporated into the KIS-CES model through the calibration of its structural parameters.

\section{Algebraic policy experiment: crisis under fiscal dominance}
\label{sec:crisis}

This section applies the Keynesian Intertemporal Synthesis (KIS-CES) framework to a critical and practical policy challenge: designing a stabilization program for an economy trapped in a state of fiscal dominance. We consider a baseline scenario characterized by a large and persistent primary deficit, where government outlays far exceed revenues:

$\left( \frac{G^{C}_0+G^{I}_0-T_0}{Y^{agg}_0}\gg 0\right)$, 

Leading to double-digit inflation $\pi_0 \gg \pi^*$ that has un-anchored expectations, and a steadily depreciating currency. In such an environment, the fiscal authority is the residual claimant on the inflation tax (seigniorage), meaning that the central bank's ability to conduct independent monetary policy is severely constrained; any attempt at monetary tightening would be politically unsustainable due to the immediate fiscal crisis it would trigger. This represents the canonical \emph{Unpleasant Monetarist Arithmetic} scenario, where fiscal needs dictate monetary outcomes. To make the analysis concrete, we calibrate the model's baseline elasticities to be consistent with the meta-analytic evidence cited throughout this paper, providing an empirically disciplined foundation for the policy experiment.

The key parameters for this experiment are the fiscal multipliers for unproductive and productive spending, the elasticity of substitution in production which underpins the multipliers, the share of liquidity-constrained households, and the responsiveness of net exports to currency movements. We set these as follows: $\{m_{G^{C}}, m_{G^{I}}\}=\{0.9, 1.6\}$, $\sigma_{prod}=0.6$, $\lambda=0.45$, and $\eta\equiv\frac{\partial NX}{\partial q}=0.15$, where $q$ is the (log) real exchange rate, such that an increase represents a \emph{devaluation}. The higher multiplier for public investment, $m_{G^{I}}$, implicitly captures both the direct demand stimulus and the  crowding-in effect on private investment ($\epsilon_{I,K^P} > 0$) that arises from the CES production structure with factor complementarity ($\sigma_{prod} < 1$). For notation, a "hat" over a variable will denote a \emph{log-linearised} percentage deviation from the initial crisis baseline, e.g., $\widehat{Y}\equiv dY^{agg}/Y^{agg}_0$.

\subsection{Generic impact multipliers}

For any small combination of policy impulses, the total short-run impact on aggregate output can be captured by a single, comprehensive linear equation derived from the model's structure. This equation serves as our analytical workhorse for all subsequent scenarios. The change in output is a weighted sum of the changes in government consumption, government investment, the real exchange rate, and the stock of public debt.
\begin{align}
\widehat{Y}
&=
m_{G^{C}}\bigl(\widehat{G}^{C}\bigr)
+
m_{G^{I}}\bigl(\widehat{G}^{I}\bigr)
+
\eta\,\widehat{q}
-
\chi\;\widehat{B},
\label{eq:generic_mult}
\end{align}
where the first two terms represent the direct fiscal impulses, weighted by their respective, and different, multipliers. The third term, $\eta\,\widehat{q}$, captures the expansionary effect of a real devaluation ($\widehat{q}>0$) on net exports. The final term, $-\chi\;\widehat{B}$, represents the \emph{wealth channel} on aggregate demand, where $\chi > 0$. When the government issues new debt ($\widehat{B}>0$), it increases the net asset position of the Wealth-Accumulating (WA) households. According to the Modified Euler Equation (where wealth is an argument in the utility function), a higher stock of assets increases the households' sense of security, allowing them to save less out of current income. However, the standard neoclassical channel suggests that higher debt implies higher future taxes, which could reduce consumption. The meta-analysis by Stanley \cite{stanley1998} finds a strong rejection of full Ricardian Equivalence, suggesting that the net effect of debt on consumption is negative but not one-for-one. We therefore model this as a net dampening effect, where $\chi\approx 0.2$, meaning a 1\% of GDP increase in debt reduces aggregate demand by 0.2\%. In scenarios with no new net debt issuance ($\widehat{B}=0$), this term vanishes.

\subsection{Strategy A: Spending shock austerity}
This strategy represents the orthodox "shock therapy" approach. It involves a one-shot, aggressive cut of unproductive government outlays, $\Delta G^{C}<0$, equivalent to $5\%$ of baseline GDP. This implies a policy shock of $\widehat{G}^{C}=-0.05$. Productive investment is left unchanged, so $\widehat{G}^{I}=0$. We explore three variants of how this austerity program is financed and supported.

\paragraph{A1. Austerity \textbf{with} new debt and devaluation.}
The fiscal authority implements the spending cut but still needs to finance a portion of the remaining deficit with new debt, equivalent to $3\%$ of GDP ($\widehat{B}=+0.03$). Simultaneously, it engineers a $10\%$ real depreciation ($\widehat{q}=+0.10$) to support the external sector. Substituting these values into our generic impact equation gives the total effect on output.
$$
\widehat{Y}_{\text{A1}}
=
m_{G^{C}}(-0.05) + m_{G^{I}}(0) + \eta(0.10) - \chi(0.03)
$$

$$
=
0.9(-0.05) + 0.15(0.10) - 0.2(0.03)
=
-0.045 + 0.015 - 0.006 = -0.036.
$$

The result is a deep economic contraction of $3.6\%$. This calculation reveals the anatomy of the shock: the direct austerity induces a $4.5\%$ fall in GDP, which is only partially cushioned by a $1.5\%$ boost from net exports. The wealth effect from new debt adds a further, albeit smaller, contractionary impulse. The conclusion is stark: even with a significant devaluation, the aggressive spending cut plunges the economy into a severe recession.

\paragraph{A2. Austerity \textbf{without} new debt.}
In this variant, the consolidation is so effective that no new debt is issued, so $\widehat{B}=0$. The exchange rate policy remains the same ($\widehat{q}=+0.10$). The impact on output is a direct simplification of the previous case.
\[
\widehat{Y}_{\text{A2}}
=
0.9(-0.05) + 0.15(0.10) = -0.045 + 0.015 = -0.030.
\]
The resulting recession is still severe, at $3.0\%$, but it is less deep than in the first scenario. The reason is that eliminating new debt issuance removes the adverse wealth channel, thereby providing a small amount of relief. This illustrates that, under non-Ricardian assumptions, the method of financing the residual deficit matters for the real economy.

\paragraph{A3. Austerity without devaluation.}
Here, we consider the harshest scenario where the government maintains a fixed real exchange rate ($\widehat{q}=0$) while implementing the austerity cut and issuing new debt ($\widehat{B}=+0.03$).
\[
\widehat{Y}_{\text{A3}}
=
0.9(-0.05) - 0.2(0.03) = -0.045 - 0.006 = -0.051.
\]
This policy mix produces the most devastating outcome: a $5.1\%$ contraction in GDP. By forgoing the cushioning effect of a devaluation, the full force of the fiscal consolidation hits the domestic economy, compounded by the negative wealth effect of rising debt. This case highlights the perilous nature of "internal devaluation" strategies, which rely solely on depressing domestic demand to achieve adjustment.

\subsection{Strategy B: Gradualism}
This strategy avoids the shock of immediate, deep cuts. Instead, it defines a three-year adjustment path to achieve the same cumulative $5\%$ of GDP reduction in unproductive spending, $G^{C}$. This means the policy impulse in the first year is only one-third of the shock therapy, or $\widehat{G}^{C} \approx -0.017$.

\paragraph{B1. Gradualism with debt and devaluation.}
We assume the auxiliary policies are also scaled down proportionally for the first year: new debt issuance is $\widehat{B}=+0.01$ and the devaluation is $\widehat{q}=+0.033$.
\[
\widehat{Y}_{\text{B1}}
=
0.9(-0.017) + 0.15(0.033) - 0.2(0.01)
=
-0.0153 + 0.005 - 0.002 = -0.0123.
\]
The first-year impact is a much milder recession of $1.2\%$. The smaller initial shock allows forward-looking agents (the WA households) and firms to better smooth their consumption and investment paths, mitigating the downturn. However, the economy still contracts, and this strategy introduces the risk that the multi-year plan may lack credibility.

\paragraph{B2. Gradualism without new debt.}
If the smaller cut is sufficient to stabilize the budget without new debt in the first year ($\widehat{B}=0$), the outcome is:
\[
\widehat{Y}_{\text{B2}}
=
0.9(-0.017) + 0.15(0.033) = -0.0153 + 0.005 = -0.0103.
\]
The recession is slightly milder, at $1.0\%$. The key insight from comparing the gradualist scenarios is that while they soften the immediate blow, they do not change the fundamental direction of the economy, which remains contractionary as long as the policy focuses exclusively on spending cuts.

\subsection{Strategy C: Aggressive spending restructuring}
This strategy represents a paradigm shift, moving the focus from the \emph{level} of spending to its \emph{composition}. The government reallocates a significant portion of the budget, $4\%$ of GDP, from low-multiplier unproductive consumption to high-multiplier productive public investment. This is a budget-neutral switch in terms of total outlays: $\widehat{G}^{C}=-0.04$ and $\widehat{G}^{I}=+0.04$.

\paragraph{C1. Restructuring with debt and devaluation.}
The government undertakes the ambitious restructuring while securing temporary bridge-financing equivalent to $1\%$ of GDP ($\widehat{B}=+0.01$) and engineering a modest $5\%$ real depreciation ($\widehat{q}=+0.05$). The impact is calculated using the full generic equation.
\[
\widehat{Y}_{\text{C1}}
=
m_{G^{C}}(-0.04) + m_{G^{I}}(+0.04) + \eta(0.05) - \chi(0.01)
=
0.9(-0.04) + 1.6(+0.04) + 0.15(0.05) - 0.2(0.01)
\]
\[
= -0.036 + 0.064 + 0.0075 - 0.002 = +0.0335.
\]
The outcome is a robust economic expansion of $3.4\%$. This result is the core prediction of the KIS-CES framework in action. Even with no change in total government spending, the shift from low-return to high-return expenditure creates a powerful net positive stimulus. The $6.4\%$ boost from productive investment, which includes the crowding-in of private capital, overwhelmingly dominates the $3.6\%$ drag from cutting consumption. The devaluation adds a further positive impulse.

\paragraph{C2. Restructuring without debt and no devaluation.}
This is the purest test of the restructuring strategy, with no auxiliary policies: $\widehat{B}=0$ and $\widehat{q}=0$.
\[
\widehat{Y}_{\text{C2}}
=
0.9(-0.04) + 1.6(+0.04) = -0.036 + 0.064 = +0.028.
\]
Even in isolation, the expenditure switch is decisively expansionary, boosting GDP by $2.8\%$. This demonstrates that the primary driver of success is the improvement in the quality of the fiscal stance, a channel that is fundamentally internal to the government's budget.

\paragraph{C3. Restructuring without debt but with devaluation.}
Finally, we combine the pure restructuring with the supportive devaluation ($\widehat{q}=+0.05$).
\[
\widehat{Y}_{\text{C3}}
=
(+0.028) + \eta(0.05) = +0.028 + 0.15(0.05) = +0.028 + 0.0075 = +0.0355.
\]
The expansion is even stronger, at $3.6\%$. This policy package appears to be one of the most effective, as it combines the powerful internal restructuring with a supportive external stimulus, all without increasing the public debt burden.

The results of these policy experiments, summarized in Table~\ref{tab:summary_crisis}, provide a clear and compelling narrative about fiscal stabilization. The numerical outcomes, all derived from the same empirically-grounded theoretical model, allow for a direct and rigorous comparison of the strategies, leading to several key lessons for policymakers.

\begin{table}[h]
\centering
\begin{tabular}{lccc}
\toprule
\textbf{Scenario} & $\widehat{G^{C}}$ & $\widehat{G^{I}}$ & $\widehat{Y}$ \\ \midrule
A1 Aggressive (debt + deval.) & $-0.05$ & $0$ & $-0.036$ \\
A2 Aggressive (no debt)         & $-0.05$ & $0$ & $-0.030$ \\
A3 Aggressive (no deval.)       & $-0.05$ & $0$ & $-0.051$ \\ \addlinespace
B1 Gradual (debt + deval.)    & $-0.017$ & $0$ & $-0.012$ \\
B2 Gradual (no debt)            & $-0.017$ & $0$ & $-0.010$ \\ \addlinespace
C1 Switch (debt + deval.)     & $-0.04$ & $+0.04$ & $+0.034$ \\
C2 Switch (no debt, no deval.) & $-0.04$ & $+0.04$ & $+0.028$ \\
C3 Switch (no debt, deval.)    & $-0.04$ & $+0.04$ & $+0.036$ \\ \bottomrule
\end{tabular}
\caption{One-period output effects under alternative stabilization strategies. Negative numbers denote recessions; positive numbers denote expansions.}
\label{tab:summary_crisis}
\end{table}

\paragraph{Key Lessons.}

\textbf{Composition trumps size.} The most striking conclusion is that the structure of fiscal adjustment is far more important than its aggregate size. A budget-neutral switch from low-multiplier consumption to high-multiplier investment is powerfully expansionary, while a simple cut in spending of the same magnitude is deeply recessionary. This highlights the critical importance of treating public investment as a distinct, productive factor.

\textbf{Debt issuance has real, non-neutral effects.} The financing of the deficit matters. Issuing new debt has a dampening effect on aggregate demand through the wealth channel, confirming the empirical rejection of strict Ricardian neutrality. Policies that can achieve stabilization without accumulating more debt are therefore preferable from a growth perspective.

\textbf{Devaluation is a valuable but partial cushion.} Currency devaluation provides a tangible boost to output by stimulating net exports. However, for empirically reasonable trade elasticities, its effect is not large enough to single-handedly offset the recessionary force of a large austerity shock. Its role is most effective when it complements an already-expansionary strategy, such as expenditure restructuring.

\textbf{Gradualism softens, but cannot reverse, the pain of austerity.} Spreading spending cuts over time reduces the magnitude of the initial recession, but it does not change the sign of the impact. The economy still contracts. Gradualism is a strategy for managing the pain of a recession, not for avoiding it. Only a change in the composition of spending can turn a fiscal adjustment from a contractionary to an expansionary event.

The KIS-CES model features state-dependent multipliers that rise in deep recessions. This implies that the qualitative ranking observed here would become even more pronounced in reality. A front-loaded austerity shock (Strategy A) would deepen the slump, which in turn would make the fiscal multiplier even larger, creating a vicious cycle. Conversely, a growth-oriented restructuring (Strategy C) would narrow the output gap, crowding-in private investment and accelerating the recovery in a virtuous cycle. The model thus provides a rigorous theoretical foundation for advocating a shift in policy focus away from the simple arithmetic of spending cuts and towards the economic logic of productive investment.

\section{Conclusion}

This paper has introduced the Keynesian Intertemporal Synthesis model with CES production (KIS-CES), an effort to build a macroeconomic framework that overcomes the theoretical fragmentation in the study of fiscal policy and is firmly anchored in the quantitative findings of the modern empirical literature, particularly those consolidated through meta-analysis. By selectively and justifiably integrating the most powerful insights from the New Keynesian and Post-Keynesian traditions, and by taking the evidence from multiple meta-analyses as a binding constraint, the KIS-CES offers a unified narrative that aligns remarkably well with a broad spectrum of the evidence.

The model has shown that it is not necessary to choose between intertemporal optimization and the relevance of effective demand, or between nominal rigidities and the importance of inequality and financial structure. It is possible to build a framework where households are optimizers, but their optimization is informed by a richer psychology that values financial security and suffers from debt stress; where monetary policy is important, but its power is attenuated by these same behaviors; and where fiscal policy regains its status as a potent and necessary stabilization tool. The effectiveness of this tool is not a universal parameter, but an endogenous variable that depends on the state of the business cycle, the composition of the fiscal impulse, and the structural characteristics of the economy, such as the distribution of income and production technology.

The central contribution of this work has been the incorporation of a CES production function with an elasticity of substitution less than unity, an imperative dictated by the meta-analysis of Gechert et al. \cite{gechert2021}. This modification, far from being a mere technical adjustment, has revealed itself to be a transformative element. It has allowed the model to generate a crowding-in mechanism for private investment in response to public investment, providing a robust microfoundation for the empirical evidence (Gechert, 2015; Bom and Ligthart, 2014) that investment in infrastructure is the most potent form of fiscal stimulus. Together with its other features—a high and heterogeneous MPC that invalidates Ricardian Equivalence (Stanley, 1998) and a multiplier that is magnified at the ZLB (Gechert and Rannenberg, 2018)—the KIS-CES emerges as a theoretically coherent and empirically disciplined framework.

The avenues for future research are numerous. The model could be extended to include an explicit financial sector and model Minskyan instability, a task that the survey by Nikolaidi and Stockhammer \cite{nikolaidi2017} shows to be an active field. It could also be adapted to an open economy to analyze the complex phenomena of balance-of-payments crises, transfer problems, and balance sheet issues that Krugman \cite{krugman1999} identified as central to emerging market crises. The interaction between sovereign default and business cycles, modeled by Mendoza and Yue \cite{mendoza2012}, could be integrated to analyze how fiscal policy is conditioned by country risk. However, even in its current form, the KIS-CES provides a robust and empirically disciplined platform for rethinking the macroeconomics of fiscal policy. Its central message is clear: fiscal policy works, but how and how much it works depends on the context. To ignore this state-dependency and the heterogeneity of instruments is to miss the most important lessons of the last decade of economic research.


\begin{thebibliography}{99}
\bibitem{blanchard2013} Blanchard, O., \& Leigh, D. (2013). Growth forecast errors and fiscal multipliers. \textit{American economic review}, 103(3), 117-20.
\bibitem{bom2014} Bom, P. R. D., \& Ligthart, J. E. (2014). What have we learned from three decades of research on the productivity of public capital?. \textit{Journal of economic surveys}, 28(5), 889-916.
\bibitem{bonanno2024} Bonanno, G., Errico, L., Fiorino, N., \& Ricciuti, R. (2024). The impact of government size on corruption: a meta-regression analysis. \textit{SSRN Electronic Journal}.
\bibitem{bumann2013} Bumann, S., Hermes, N., \& Lensink, R. (2013). Financial liberalization and economic growth: a meta-analysis. \textit{Journal of international money and finance}, 33, 255-281.
\bibitem{christiano2011} Christiano, L. J., Eichenbaum, M., \& Rebelo, S. (2011). When is the government spending multiplier large?. \textit{Journal of political economy}, 119(1), 78-121.
\bibitem{churchill2016} Churchill, S. A., Ugur, M., \& Yew, S. L. (2016). Does government size affect per-capita income growth? a hierarchical meta-regression analysis. \textit{Economic record}, 92(299), 551-571.
\bibitem{flood1984} Flood, R. P., \& Garber, P. M. (1984). Collapsing exchange-rate regimes: some linear examples. \textit{Journal of international economics}, 17(1-2), 1-13.
\bibitem{garcia-cicco2010} García-Cicco, J., Pancrazi, R., \& Uribe, M. (2010). Real business cycles in emerging countries?. \textit{American economic review}, 100(5), 2510-31.
\bibitem{gechert2015} Gechert, S. (2015). What fiscal policy is most effective? a meta-regression analysis. \textit{Oxford economic papers}, 67(3), 553-580.
\bibitem{gechert2018} Gechert, S., \& Rannenberg, A. (2018). Which fiscal multipliers are regime-dependent? a meta-regression analysis. \textit{Journal of economic surveys}, 32(4), 1160-1182.
\bibitem{gechert2021} Gechert, S., Havranek, T., Irsova, Z., \& Kolcunova, D. (2022). Measuring capital-labor substitution: the importance of method choices and publication bias. \textit{Review of economic dynamics}, 45, 55-82.
\bibitem{gechert2022a} Gechert, S. (2022). Reconsidering macroeconomic policy prescriptions with meta-analysis. \textit{Industrial and corporate change}, 31(2), 576-590.
\bibitem{gechert2022b} Gechert, S., \& Heimberger, P. (2022). Do corporate tax cuts boost economic growth?. \textit{European economic review}, 147, 104157.
\bibitem{gechert2023a} Gechert, S. (2023). Fiscal policy: post- or new keynesian?. \textit{European journal of economics and economic policies: intervention}, 20(2), 338-355.
\bibitem{heimberger2023} Heimberger, P. (2023). The cyclical behaviour of fiscal policy: a meta-analysis. \textit{Economic modelling}, 123, 106259.
\bibitem{irsova2024} Irsova, Z., Doucouliagos, H., Havranek, T., \& Stanley, T. D. (2024). Meta-analysis of social science research: a practitioner's guide. \textit{Journal of economic surveys}, 38(5), 1547-1566.
\bibitem{kaminsky1999} Kaminsky, G. L., \& Reinhart, C. M. (1999). The twin crises: the causes of banking and balance-of-payments problems. \textit{American economic review}, 89(3), 473-500.
\bibitem{karlan2025} Crosta, T., Karlan, D., Ong, F., Rüschenpöhler, J., \& Udry, C. R. (2025). Unconditional cash transfers: a bayesian meta-analysis of randomized evaluations in low and middle income countries. \textit{NBER Working Paper Series}, No. 32779.
\bibitem{keynes1936} Keynes, J. M. (1936). \textit{The general theory of employment, interest and money}. London: Macmillan.
\bibitem{krugman1999} Krugman, P. (1999). Balance sheets, the transfer problem, and financial crises. In \textit{International finance and financial crises: essays in honor of robert p. flood, jr.} (pp. 31-55). Springer, Dordrecht.
\bibitem{mendoza2012} Mendoza, E. G., \& Yue, V. Z. (2012). A general equilibrium model of sovereign default and business cycles. \textit{The quarterly journal of economics}, 127(2), 889-946.
\bibitem{nikolaidi2017} Nikolaidi, M., \& Stockhammer, E. (2017). Minsky models: a structured survey. \textit{Journal of economic surveys}, 31(5), 1304-1339.
\bibitem{nijkamp2004} Nijkamp, P., \& Poot, J. (2004). Meta-analysis of the effect of fiscal policies on long-run growth. \textit{European journal of political economy}, 20(1), 91-124.
\bibitem{stanley1998} Stanley, T. D. (1998). New wine in old bottles: a meta-analysis of ricardian equivalence. \textit{Southern economic journal}, 64(3), 713-727.
\end{thebibliography}
\end{document}